# First Radar – CubeSat Transionospheric HF Propagation Observations: Suomi 100 Satellite and EISCAT HF Facility

Esa Kallio[1], Antti Kero[2], Ari-Matti Harri[3], Antti Kestilä[3], Anita Aikio[2], Mathias Fontell[1], Riku Jarvinen[1,3], Kirsti Kauristie[3], Olli Knuuttila[1], Petri Koskimaa[3], Jauaries Loyala[1], Juha-Matti Lukkari[1], Amin Modabberian[1], Joonas Niittyniemi[1], Jouni Rynö[3], Heikki Vanhamäki[2], and Erik Varberg[4]

[1]*Aalto University, School of Electrical Engineering, Espoo, Finland*

[2]*University of Oulu, Oulu, Finland*

[3]*Finnish Meteorological Institute, Helsinki, Finland*

[4]*EISCAT Scientific Association, Ramfjordmoen, Norway; UiT The Arctic University of Norway, Tromsø, Norway*

*Corresponding author: Esa Kallio (esa.kallio@aalto.fi)*

**Key Points:**

- Active radar-satellite transionospheric measurements in the HF range with a CubeSat-size satellite have been performed.

- A radio spectrometer onboard the Suomi 100 CubeSat detected a clear signal transmitted by the EISCAT HF heater.

- The properties of ionospheric plasma were found to be affected by the energy of the transmitted heater signal.

**Abstract**

Radio waves provide a useful diagnostic tool to investigate the properties of the ionosphere because the ionosphere affects the transmission and properties of High Frequency (HF) electromagnetic waves. We have conducted a transionospheric HF-propagation research campaign with a nanosatellite on a low-Earth polar orbit and the EISCAT HF transmitter facility in Tromsø, Norway, in December 2020. In the active measurement, the EISCAT HF facility transmitted sinusoidal 7.953 MHz signal which was received with the HEARER radio spectrometer onboard 1 Unit (size: 10 cm × 10 cm × 10 cm) Suomi 100 space weather nanosatellite. Data analysis showed that the EISCAT HF signal was detected with the satellite's radio spectrometer when the satellite was the closest to the heater along its orbit. Part of the observed variations seen in the signal was identified to be related to the heater's antenna pattern and to the transmitted pulse shapes. Other observed variations can be related to the spatial and temporal variations of the ionosphere and its different responses to the used transmission frequencies and to the transmitted O- and X-wave modes. Some trends in the observed signal may also be associated to changes in the properties of ionospheric plasma resulting from the heater's electromagnetic wave energy. This paper is, to authors' best knowledge, the first observation of this kind of "self-absorption" measured from the transionospheric signal path from a powerful radio source on the ground to the satellite-borne receiver.

**1 Introduction**

Numerous different ground-based research instruments use radio waves to investigate the properties of the ionosphere. Riometers are receivers which measure the opacity of the ionosphere to natural radio signals coming from space. An ionosonde includes both a transmitter and a receiver and it can provide information about the critical frequencies of the ionospheric layers and their virtual heights. Incoherent scatter radars require very high power and they yield information about the altitude profiles of several plasma parameters.

Moreover, investigation of the properties of the ionosphere and space above it with the aid of satellites has been known for a long time to provide new possibilities because a satellite can measure the top-side ionosphere above the peak electron density (e.g., King, 1963; Franklin and Maclean, 1969; Rothkaehl and Klos, 1996; Karpachev et al., 2012). In particular, radio instruments onboard satellites orbiting above the peak electron density of the ionosphere can measure low radio frequency (RF) waves which cannot propagate through the ionosphere and, consequently, cannot be observed from the ground. Therefore, such satellite-based radio instruments can be used to investigate natural HF waves from our Solar System, such as type II solar radio bursts (see e.g., Winter L. M. and K. Ledbetter, 2015), and beyond. Conversely, satellites can receive natural and man-made radio signals which originate from the ground if their frequency is sufficiently high to be transmitted through the ionosphere. In addition to Earth's ionospheric research, these methods have also been used to investigate the Martian ionosphere with the MARSIS radar onboard the Mars Express orbiter (see e.g., Kopf et al., 2008).

Medium Frequency (MF: 0.3 – 3 MH) and High Frequency (HF: 3 – 30 MHz) observations have been made by many large satellites. For example, ISIS-I and ISIS-II spacecraft launched in 1978, made RF measurements in the frequency range of 0.1 – 20 MHz (Franklin and Maclean, 1969). Later, for example, Intercosmos-19 (1979 – 1981), Intercosmos-24 (launched in 1989), Intercosmos-25/APEX (launched in 1992) and Coronas-I (1994 – 2001) continued RF measurements in various frequency ranges: Intercosmos-19 in ~ 0.6 – 16 MHz (Karpachev et al., 2013), Intercosmos-24 in ~ 0.1 – 10 MHz (Teodossiev et al., 2001), Intercosmos-25/APEX in ~ 0.1 – 20 MHz (Dokukin, 1992; Prech et al., 2018) and Coronas-I in 0.1 – 30 MHz (Oraevsky et al., 1998). Later, the ALEXIS spacecraft (1993 – 2005) made high resolution VHF measurements of the electric field from about 25 MHz to 100 MHz with the use of its BLACKBEARD instrument (Holden et al., 1995). Later, the Fast on-Orbit Recording of Transient Events (FORTE) spacecraft (launched 1997) measured the 20 – 300 MHz frequency range (Massey et al., 1998). The more recent HF measurements have been made with the e-POP instrument on the CASSIOPE spacecraft which measured the electric field using the bandwidth from 10 Hz to 18 MHz (James et al., 2015).

One specific application for the satellite-borne radio instruments, which was investigated already in the earlier HF-satellite receives, is to detect signals sent from a ground transmitter station. The obvious advantage of such joint satellite and ground-based measurement is that interpretation of the measurements is simplified when the properties of the signal and the position of the signal source are known in detail. For example, in 1978, a transmitter located at Ottawa sent a 9.303 MHz signal to the ISIS-I and ISIS-II spacecrafts over 100 times (James, 2006). In 1981, the DE 1 satellite observed a signal where a 2.759 MHz heater carrier, which was amplitude modulated with an ELF/VLF wave, was transmitted from the heater facility near Tromsø, Norway (James et al., 1990). Later, the FORTE spacecraft was used in the active measurement when it observed fast, about a nanosecond in lenght, signals transmitted from the electromagnetic pulse generator (LAPP), Los Alamos, New Mexico (Massey et al., 1998). Moreover, the Radio Receiver Instrument (RRI) onboard the e-POP spacecraft has been used to make joint measurements with the SPEAR ionospheric heater at Longyearbyen, Svalbard, Norway, where it observed the transmitted 4.45 MHz signal (James et al., 2015), and the EISCAT ionospheric heater's 3.9 MHz and 5.423 MHz signals (Leyser et al., 2018). The RRI has also observed a 11.2 MHz signal transmitted with the Super Dual Auroral Radar Network (SuperDARN) at Saskatoon and Rankin Inlet (Perry et al., 2017). The receiver's sensitivity enabled also observing signals transmitted by radio amateurs (Perry et al., 2018). The RRI has also been used successfully to observe VLF (Very Low Frequency) and ELF (Extremely Low Frequency) signals generated by the Sura heating facility (Vas'kov et al., 1998). The Sura heater has also been used in active measurements with the APEX and the CORONAS satellites (Oraevsky et al., 1998), the DEMETER satellite (Zhang et al., 2016) and the RRI/e-POP/CASSIOPE instrument (James et al., 2017).

Although these space missions with radio instrumentation have provided unique information about the properties of the ionosphere working either alone, or jointly with ground-based instruments, the number of radio instruments in space is much less than the number of different radio instruments on the ground. One reason for this is the high cost of the satellites. The total cost of each satellite mission, in turn, depends partly on the size and the mass of the satellite. Historically, the mass of a satellite with a radio instrument has been more than a hundred kilograms. For example, the mass of the ALEXIS satellite, which is sometimes stated to be one of the first modern and sophisticated miniature satellites, weighted 113 kg (Priedhorsky et al., 1993). The mass and the size of the dedicated radio instrument can also be large, such as the RRI on e-POP, which had a mass of 8.68 kg and a volume of 7400 cm$^{-3}$ (James et al., 2015).

However, the recent miniaturizing and standardization of satellite components has made it possible to manufacture significantly lighter and cheaper satellites, the so-called CubeSats, for different space activities. In the beginning, the CubeSats were used mainly for educational purposes, then later also in technology demonstrations, and nowadays more and more for various commercial purposes. Furthermore, CubeSat technology has also generated interest in space science as a result of numerous science-focused Cubesat projects (see e.g., Poghosyan A. and Golkar, 2017; Streltsov et al., 2018, and references therein). The small cost of the CubeSat compared to conventional large satellites can also open a door for multi-point measurements, such as those performed within the Dynamic Ionosphere CubeSat Experiment (DICE) mission, consisting of two 1.5 Unit CubeSats launched in 2011 (Fish et al., 2014).

The possibilities of using a CubeSat in joint measurement with ground-based devices has been tested recently. The first step to use a nanosatellite, i.e., a satellite in the mass range of 1 – 10 kg, in joint satellite-radar VHF measurements was a 3 Unit Radio Aurora Explorer CubeSat, RAX (Bahcivan and Cutler, 2012). RAX-1 was launched in 2010 and RAX-2 in 2011. The RAX-2 satellite observed coherent scattering from the 449 MHz pulses transmitted from the Poker Flat Incoherent Scatter Radar (Bahcivan et al., 2012, 2014). These measurements indicated that active satellite-radar VHF measurements are feasible. Active measurements are also planned with a 2-Unit INSPIRE-SAT 7 which IONO experiment will observe 8 – 20 MHz signals from a ground-based HF transmitter [Meftah et al., 2022].

In this paper we present the first results of the research project, where a radio spectrometer measurement in the HF range was made with the smallest-sized CubeSat (1 Unit), i.e., with a satellite with a volume of 10 cm × 10 cm × 10 cm. The paper presents the result of the measurement campaign made in December 2020 where the Suomi 100 satellite, the first Finnish research satellite, was tuned to detect frequencies of 7.953 MHz transmitted from the EISCAT heater facility in Tromsø, Norway. According to the authors' best knowledge, the measurement was the first ever active joint CubeSat-radar measurements in the HF range. We also analyze the observed signal with a ray-tracing simulation using an empirical electron density altitude profile based on the EISCAT VHF and Tromsø digisonde measurements.

The paper is organized as follows. First, we introduce the Suomi 100 satellite mission and its payloads as well as the EISCAT heater facility. Then we describe some basic features of how active satellite-radar measurements can be used to investigate the ionosphere and use the Suomi 100 satellite as an example case. Finally, we present the satellite's measurements in Dec. 9, 2020, where the EISCAT heater signal was detected, and interpret the observations.

## 2 Instruments

### 2.1 Suomi 100 mission and its radio spectrometer

The Suomi 100 satellite (COSPAR ID 2018-099AY), 1 Unit (size: 10 cm × 10 cm × 10 cm) excluding antennas deployed from the spacecraft bus) about 1.3 kg, was launched on Dec. 3, 2018, to a circular Sun synchronous orbit of the peak altitude of 600 km and inclination of 98°. The research goal of the satellite has been to investigate the ionosphere and auroras. The satellite has two payloads, a radio spectrometer (Fig. 1) and a wide-angle white-light camera. The camera is used to photograph auroras (Knuuttila et al., 2022) while the goal of the radio spectrometer is to measure natural and man-made radio waves in the HF range. The HEARER (High frEquency rAdio spectRomEteR) instrument was designed, built, and tested at Aalto University, Finland. The narrow-band superheterodyne receiver and its antennas are optimized to observe radio waves in the frequency range of approximately 5–10 MHz.

Observing HF radio waves in space is a challenging task because of their long wavelengths. For example, the wavelength of 10 MHz radio waves in a vacuum is 30 meters. Including a single or half wavelength electrical antenna in a spacecraft is a challenging task, especially in a 1 Unit

CubeSat which has as an additional relatively large payload, a camera with a large mass (169 g). Therefore, the HEARER uses conventional ferrite coil antennas, which are mechanically much easier to implement than a long wire antenna. The instrument has two ferrite rod antennas: Antenna 1 (MFA: Medium Frequency Antenna) and Antenna 2 (HFA: High Frequency Antenna). The antennas have different inductance due to the different number of windings around the ferrite core: Antenna1 and 2 have 110 and 22 turns, respectively, so that Antenna 1 is optimized for the lower and Antenna 2 for the higher part of the frequency range (see Koskimaa, 2016, for more details about antenna design). Two 76 × 13 × 5.6 mm ferrite rods made of Fair-Rite ferrite material (material number 61, NiZn), both weighting 26.9 g, are located on both sides of the camera objective. The rods are supported by two antenna-holding structures manufactured by using the ULTEM plastic in 3D printing. The mass of the ULTEM frames is 40 g. The superheterodyne receiver was made by using a commercial off-the-shelf AM/LW/SW/FM/WB chip Si4743-C10-GM from Silicon Labs. The mass of the HEARER's electric board is 51 g. The total mass of the HEARER, including its two antennas, the ULTEM frames and the electric board, is 144.8 g.

The HEARER instrument has three measurements modes: the raw mode (RAW), the average radiometer mode (ARM) and the average radio spectrometer (ARS) mode. The RAW mode is planned to make relatively short, typically in the order of a few minutes, measurements with the maximum sampling rate of 1000 data points per second, with a single constant measurement frequency. The ARM mode also uses constant frequency but reduces the amount of the downlinked data by calculating statistical estimators over short sampling periods, e.g., averaging over 1000 data points resulting in a time resolution of one second. The five calculated statistical estimators are minimum, maximum, average and median values as well as the standard deviation. The ARS mode is similar to the ARM mode except that it enables sweeping over different frequencies.

The position of the satellite is not exactly known because it does not have a Global Navigation Satellite System (GNSS) receiver, however the position of the satellite can be derived from Two-Line Elements (TLEs).

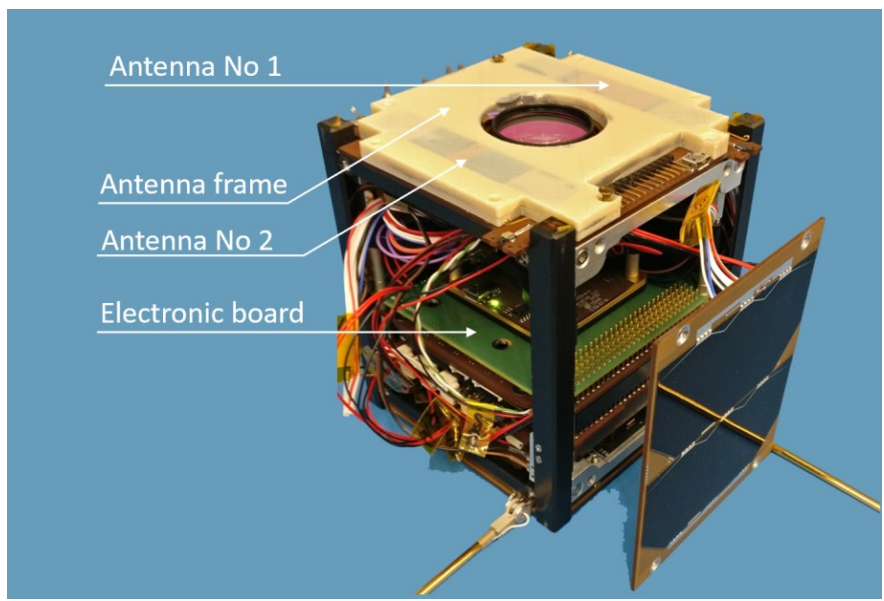

**Figure 1**. Suomi 100 satellite structure (without the solar panels on each side). The position of the HEARER radio spectrometer's electronic board, its two ferrite rod antennas, and the antenna-holding frame are marked with arrows.

2.2 EISCAT HF Heating Facility

The EISCAT HF Heating Facility is located in Tromsø, Norway (19°13'00''E, 69°35'00''N). It consists of three large phased-antenna arrays capable of transmitting radio waves in the frequency range of 5.4–8 MHz (Array 1), 3.85–5.6 MH (Array 2), and 5.4–8 MHz (Array 3) [Rietveld et al., 2016]. The effective radiated power (ERP) of Arrays 1-3 is about 600–1250 MW, 200–350 MW and 190–350 MW, respectively (see Rietveld and Stubbe, 2021, for the history of the Tromsø ionosphere heating facility). The cross-dipole antenna elements of the arrays are capable of transmitting either right-handed or left-handed circularly polarized waves, often referred to as the O-mode and the X-mode, respectively. The antenna patterns (see Fig. 2) have typically a strong main lobe, and in addition, several weaker side lobes with local minima and maxima (Zabotin et al., 2014; Rietveld, 2016). In our measurement campaigns, Array 3 was used in 2019 and Array 1 in 2020.

The power of the antenna was changed in time in order to unambiguously identify the transmitted signal from other possible signal sources, and for studying the potential nonlinearity of the ionospheric propagation. During the RAW mode measurement, the transmission power was increased quasi-logarithmically in 0.2 second intervals: 0.0–0.2 s: 0%, 0.2–0.4 s: 4.4%, 0.4–0.6 s: 10%, 0.6–0.8 s: 24%, 0.8–1.0 s: 100%, where the percentages are with respect to the maximum output power of the heater. This repeating heating cycle is referred to later as "1s/0.2s" modulation. In another modulation used in the measurement campaigns, the heater was switched between OFF and full power in 1-second intervals (1 s ON, 1 s OFF). In the third modulation used during the ARS-mode measurement in 2019, instead, the transmission power was increased linearly in 25% steps and 2-second intervals, i.e., 0–2 s: 0%, 2–4 s: 25%, 4–6 s: 50%, 6–8 s: 75%, 8–10 s: 100%. The measurement mode was kept unchanged during each of the flybys where the 1$^{st}$ campaign initially included first 6 RAW-mode measurements followed by three ARS-mode measurements. The 2$^{nd}$ campaign included 5 RAW-mode measurements.

In order to estimate the transmitted radio wave's intensity, $I$, at a given point in space and time, we combine the spatial antenna pattern (see Fig. 2) with the instantaneous transmission power like

$$I(h,\theta,\alpha,t) = \frac{G(\theta,\alpha)P_T(t)}{(4\pi h/cos\theta)^2}, \qquad (1)$$

where G is the antenna pattern as a function of the zenith angle, θ, and the azimuth angle, α, while $P_T$ is the transmitter power at time $t$, and $h$ is the altitude. This intensity would be obtained without any ionospheric effects, i.e., refraction or absorption. As can be seen in the antenna pattern of Array 1 during the flyby on Dec. 9, 2020, flyby when the O-mode was transmitted (Fig. 2), the antenna pattern includes also many local maximum values in many elevation and azimuthal directions. These spatial variations in power must be taken into account in the data analysis.

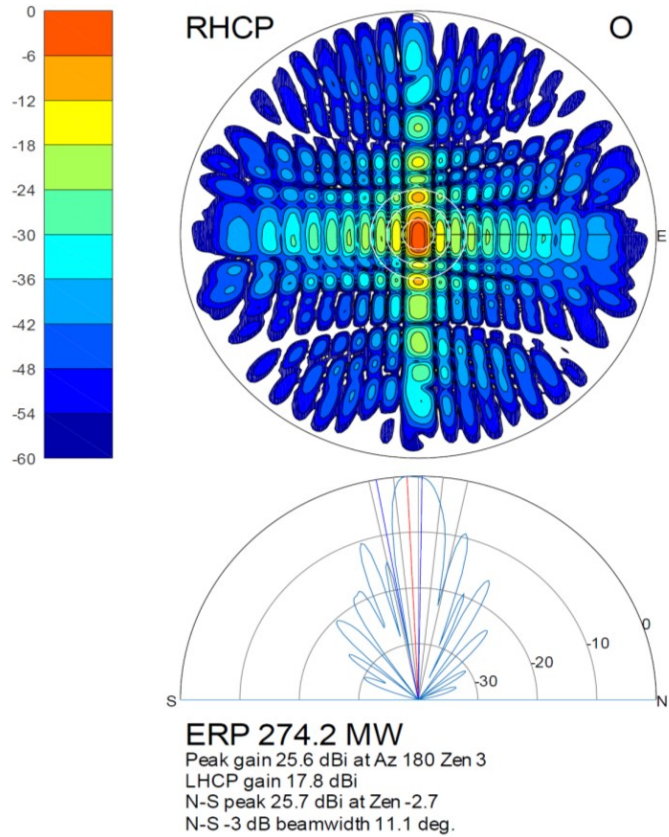

**Figure 2**. EISCAT heater Array 1's antenna pattern on Dec. 9, 2020, at 10:16:15 UT in the O-mode at 7.953 MHz when 100% transmission power (748.0 kW) was used.

## 3 Observations

Joint measurements where the HEARER measured signals from the EISCAT heater were made in two measurement campaigns: first on Jul. 15–19, 2019, and the second on Dec. 7–9, 2020. The 1$^{st}$ campaign included 9 flybys and the second 5 flybys. In three cases in 2020, the transmitted wave was the X-mode (Dec. 7. from 10:05–10:25 UT, Dec. 8. from 10:10–10:35 UT and 19:38–19:56 UT) and in all other cases O-mode waves. The EISCAT heater signal was identified in three of the five flybys during the second campaign, on Dec. 8 and 9, 2020. The clearest signal was detected during the Dec. 9, 2020 flyby, and this observation is analyzed in this work.

### 3.1 EISCAT VHF and Tromsø digisonde measurements

During the Suomi 100 campaigns, the EISCAT incoherent scatter VHF facility was used to probe the lower part of the ionosphere between altitudes of 20–197 km ionospheric electron density by using the so-called "manda" experiment mode with the antenna pointing vertically upward, which is used during low-range observations (see e.g., Tjulin, 2017). The VHF was operated for 2 hours for each flyby, starting slightly before the Suomi 100 satellite overpass. Fig. 3a shows the analyzed VHF measurements on Dec. 9, 2020, when the most intense heater signals were measured. As can be seen in Fig. 3a, the low altitude ionosphere has both fast, minute scale variations, and longer in the order of ten to several tens of minutes. Those density variations affect the propagation of the EISCAT heater signals.

The electron density profiles are investigated in more detail in Fig. 3b which show the electron densities during the time 10:20–10:35 UT when the Suomi 100 satellite made measurements. Note that there are substantial density variations at and below the D layer, the peak of which is located at an altitude of approximately 100 km. These variations are important for the propagation of HF waves because the D layer plays an important layer role in the attenuation of HF waves (see e.g., Fontell, 2018, Fig. 35, for the ray-tracing simulations), although the D layer is extremely weak at 60-90 km altitudes. The most pronounced feature is the sporadic E layer slowly descending from 105 to 100 km altitude. In Fig. 3b, the International Reference Ionosphere (IRI) model densities are also shown for a comparison. The IRI model includes density enhancement in the D-layer at ~100 km, which is qualitatively, as well as quantitatively, much as observed. However, above the D-layer the density does not increase monotonically as in the IRI model. Moreover, the measured density is higher at 200 km than predicted by the IRI model. Note that the VHF observations do not extend to the peak altitude of the electron density and, therefore, these measurements do not provide information about the peak electron density at the F layer. Therefore, the maximum observed electron density shown in Fig. 3b, the altitude of which is lower than that predicted by the IRI model, is obtained from the digisonde. However, the trend of the observations above 150 km suggests that during the Suomi 100 flyby, the electron densities were larger than predicted by the IRI model. The observations therefore suggests that analysis based on an average model, such as the IRI model, can be regarded to provide only the first-order approximation for the real electron densities during the flyby.

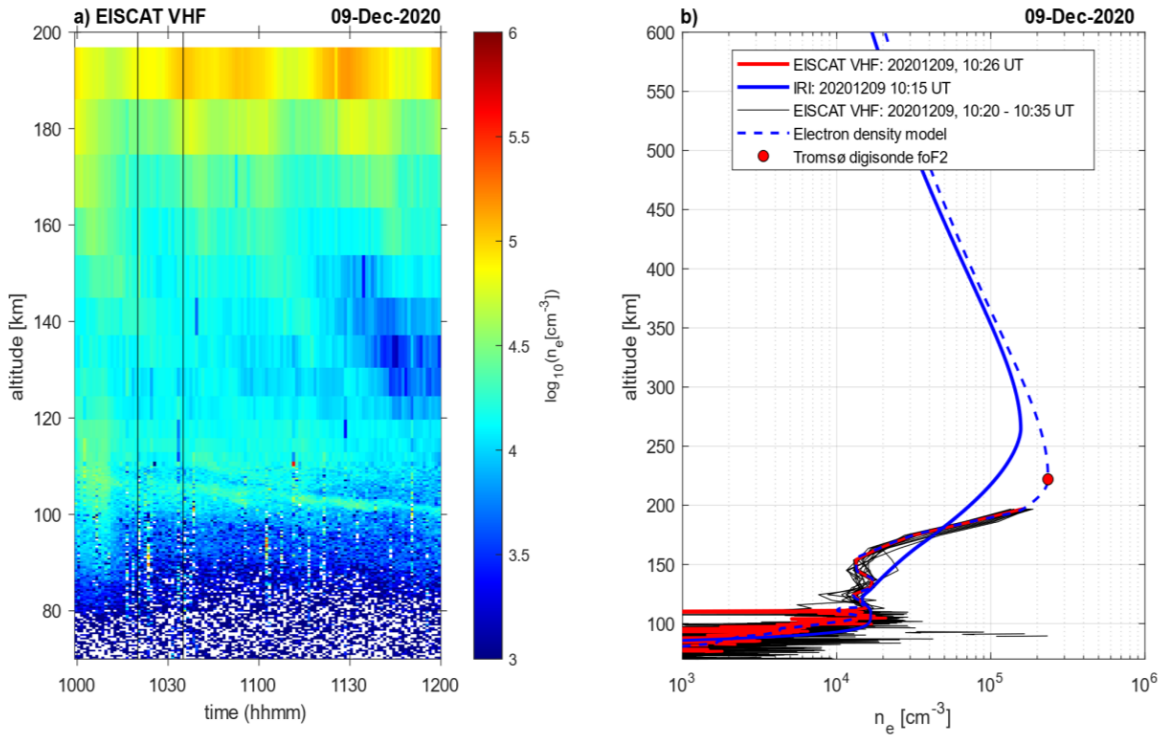

**Figure 3**. E EISCAT VHF electron density observations on Dec. 9, 2020. (a) Electron density altitude profiles during the 2-hour time when VHF was used on the Suomi 100 satellite's flyby. The region within the black vertical lines is the time range 10:20–10:35 UT when the Suomi 100 made measurements. (b) The electron density values shown in a) on 10:20–10:35 UT displayed in a density-altitude plot (black lines). The profile closest to the time 10:26 UT when the satellite was near the EISCAT heater is shown in a red line. The blue solid line shows the IRI electron density profiles which are derived at 10:15 UT. The blue dashed line shows the developed empirical density profile where the highest density is the density measured by Tromsø digisonde (the red dot). The narrow density enhancements at ~100 km is sporadic E.

3.2 Suomi 100 satellite's measurements

The orbits of the Suomi 100 satellite near the EISCAT heater on Dec. 9, 2020, during the time when the HEARER instrument observed the strongest signal, are shown in Fig. 4. The measurement was made during the whole time, when the satellite was above the horizon with respect to the EISCAT site in order to cover the full spatiotemporal variability of the strength of the detected signal, including possible leakages from the Heater side lobes. As already mentioned, the most favourable flyby took place on Dec. 9, 2020, when the satellite trajectory was slightly north from the Heater at its closest distance from the station.

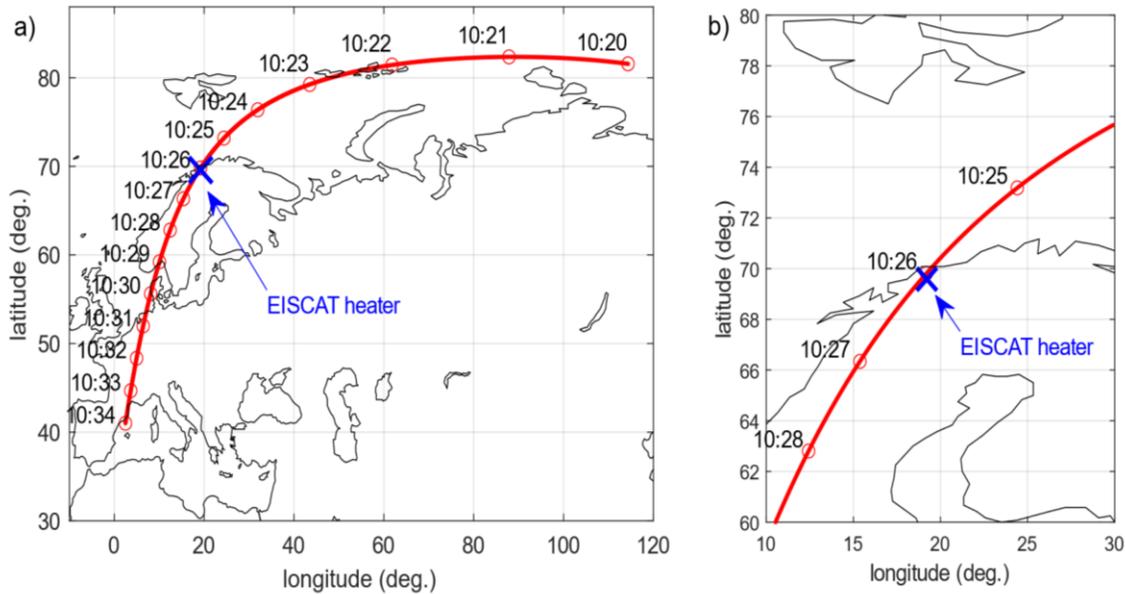

**Figure 4**. a) The positions of the Suomi 100 satellite on Dec. 9, 2020, flyby (red line). The red circles show the position of the spacecraft in 1-minute time intervals. The text shows the UT time in hh:mm. The position of the EISCAT heater is shown with a blue cross and an arrow. b) The same plot as shown in a) but is now enlarged close to the position of the EISCAT heater in order to see clearer the orbits near the transmitter.

Fig. 5 shows the HEARER's measurement on the Dec. 9, 2020, flyby near the heater. The transmitted 7.953 MHz signal was O-mode and the satellite's maximum elevation angle at the heater was 88.5°. As visible in Fig. 5 which shows the measurements near the heater, one can identify periodic pulses where the magnitude of the count increases to its maximum and then the magnitude of the count drops below the background noise. The magnitude of the count is proportional to the magnitude of the observed electric field. This kind of behavior is expected to be seen if the observed signal is the transmitted "1s/0.2s" ramp cycle signal. Note that the periodic count spikes that exceed 2000 are caused by the HEARER's data operations.

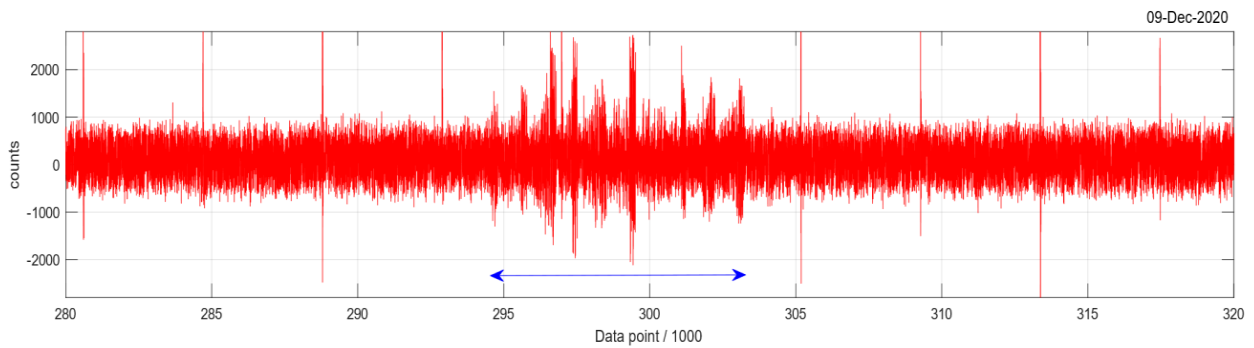

**Figure 5**. Observations of the Suomi 100 satellite on its Dec. 9, 2020, flyby, when it detected EISCAT heater's signal. The blue horizontal line shows the data regions of the measured heater signal. The value of the count shows the magnitude of the electric field, and the horizontal axis shows the number of the measured datapoint divided by 1000.

**4 Analysis of the observed EISCAT heater signal**

In this section, the basic properties of the observed radio waves are investigated by a numerical ray-tracing simulation, referred to in this publication as Aalto Ray, which was developed at Aalto University to investigate the propagation of radio waves in the ionosphere. The Aalto Ray model uses the Appleton-Hartree dispersion relation with the Hamiltonian ray equations to derive the radio ray path in a 3D space. Typically, the simulation uses a 1D electron density and the temperature profiles derived from the International Reference Ionosphere (IRI) model version 2016, and 1D neutral density atmosphere profiles from the NRLMSISE-00 atmosphere model (see Fontell, 2019, for details of the used ray-tracing simulation and its inputs). However, also manually given electron density profiles can be used.

The situation on Dec. 9, 2020, is investigated with the ray-tracing simulation in Fig. 6 which shows the ray paths of the 7.953 MHz signal which was launched at the location of the EISCAT heater. The used electron density profile was the empirically-made profile shown in Fig. 3b (the blue hashed line). All other necessary parameters, neutral densities and electron temperatures, were adopted from the IRI and NRLMSISE-00 models. Rays were transmitted from the Heater to nine elevation angles from 10º to 90º in 10º angle steps and to eight azimuth angles from 0º to 315º in 45º angle steps.

As can be seen in Fig. 6, rays launched at an elevation angle of 10º bounce from the ionosphere back to the ground. Rays launched at an elevation angle of 20º or larger, instead, propagate through the ionosphere and can reach the spacecraft altitude. It is worth noting that the rays transmitted at the elevation angle of 80º form relatively straight lines and a cone-like structure above the transmitter. The similar geometry of rays near the Heater suggests that a nominal ionosphere above the Heater may not affect significantly the signal that a satellite can observe above the Heater. Therefore, the possible variations in the observations are associated with the Heater and its ionospheric effects.

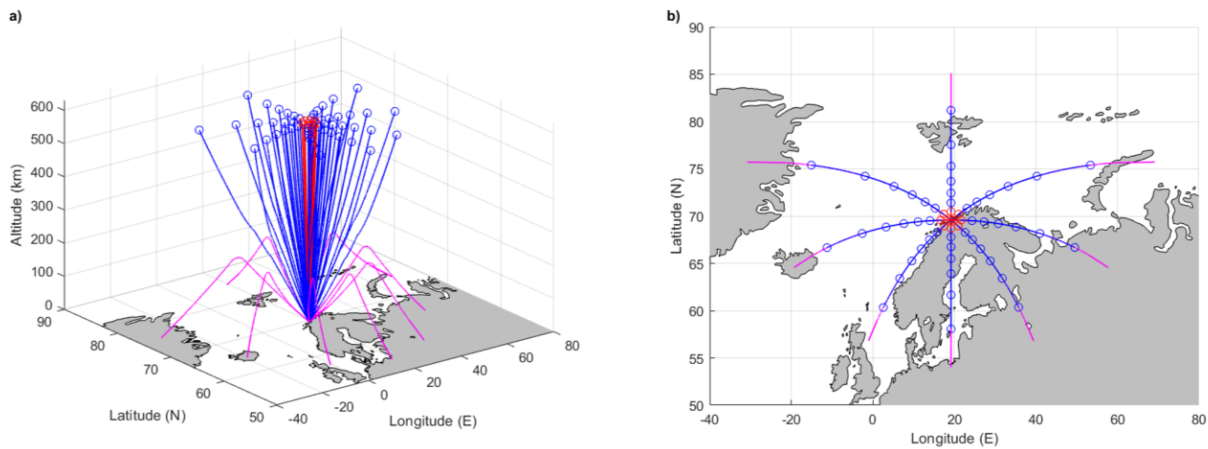

**Figure 6**. Ray paths made for the flyby on Dec. 9, 2020. Ray paths from the EISCAT Heater's 7.953 MHz O-mode waves launched from the position of the Heater. Magenta lines show rays which are bounced back to the ground, while blue and red lines identify the rays which go into space. Rays with an elevation angle of 80º and 90º are marked in red color. The position where a ray reached the altitude of 600 km is shown by a circle. Panel a) shows a side view of the 3D rays, while in b) the view direction is above the Heater. The ray tracing was made for 10:26:05 UT, when the Suomi 100 satellite was closest to the EISCAT heater.

Ray-tracing simulation can also give information of the attenuation caused by electron-neutral collisions. This attenuation along a signal ray path was investigated by calculating attenuation along a ray with an elevation angle of 90º from the heater, i.e., directly upward from the Heater. According to the developed empirical electron density model show in Fig. 3b (the dashed blue line), the attenuation of the intensity of the wave, $I$[W/m$^2$], during the flyby on Dec. 9, 2020, for the used frequency of 7.953 MHz was only ~1.3%. However, this attenuation calculation does not take into account a possible "self-absorption" effects due to the powerful HF wave's capability of heating the electron gas along the propagation path and consequently change the opacity of the plasma to the radio waves.

In Fig. 7, the observed counts from the Dec. 9, 2020, observations are compared with the trend of the electric field of the transmitted EISCAT heater signal. As already noted, during the EISCAT's "1s/0.2 s" transmission mode, the power was increased as 0%, 4.4%, 10%, 24% and 100% of its maximum power during one second period. Therefore, because the magnitude of the electric field ($E_{EISCAT}$) of the transmitted signal is proportional to the square root of the transmitted power $P_{EISCAT}$ (~ $E^2_{EISCAT}$), the electric field increases during one cycle as $\sqrt{0.044}$(~0.21), $\sqrt{0.1}$(~0.32), $\sqrt{0.24}$(~0.49), and 1 of its maximum value. In the plot, the time gap at every 4096 data blocks was assumed to be 280 ms. In addition, analysis of the timing of the RAW-mode measurements has indicated that there exists a time delay in the order of several tens of seconds from the time when the HEARER instrument was commanded to be on and when the 1$^{st}$ data point is actually saved. In Fig. 7, the initial time delay was assumed to be 43 seconds because then the location of the peak signal matches best to the EISCAT Heater modulation, as can be seen later when the observed signal and simulated EISCAT heater signal strength with the antenna pattern are compared.

In Fig. 7a and 7b, the observed signal was compared with a synthetic EISCAT's "1s/0.2 s" intensity shape, where the peak of the simulated signal was adjusted close to the observed maximum count rate. The attenuation of the signal was assumed to be unchanged during the analyzed time periods of approximately 12 seconds because the strength of the signal does not change noticeably over such a short time period around the closest distance from the Heater. As can be seen in 12 one-second time periods, the signal appears at the time period No 2, reaches its highest values at the 4$^{th}$, 5$^{th}$ and 7$^{th}$ time period, and the final signal is observed during the 11$^{th}$ time period. Although the background noise and statistical fluctuations make it difficult to make a very detailed comparison between the observed and simulated data, one can see in Fig. 7a from the observed counts a similar type of intensity increase followed by a sudden decrease of the count, as expected to be seen in the signal from the EISCAT heater.

In Fig. 7b, the observed signal was also analyzed by calculating 200 data point, i.e., 0.2 s running mean and running maximum 200 points. This means that the center running mean points at every power step contain an observation made during the same transmission power value and, consequently, are the "cleanest" mean values. Other running mean points, instead, contain an observation collected during two transmission power times. The running mean and running maximum modify the shape of the initial data and, therefore, in Fig. 8b, the simulated EISCAT heater signal is also treated in a similar way by calculating its 200-point maximum. If the signal is from the EISCAT heater, one would anticipate the running mean to increase, when the transmitted power is increased from 0% to 100% of its maximum.

As can be seen in Figs. 7a and 7b, the observed signal, the derived running mean and, especially, the derived running maximum, have similar periodic behavior as the transmitted signal. Clear differences between the presented 12 periods can, however, also be identified. In period Nos 1–5 and 10, the running max values have a clear practically monotonic increase of the signal from the minimum to the maximum value. Note that the reason to show the EISCAT power steps together with the observations in Fig. 7a is to illustrate that the periodic changes in the data are correlated with the periodic changes of the transmitted signal without taking at this stage into account EISCAT's antenna pattern. However, one can see that, for example, on period Nos 4 and 5, where the maximum value of the illustrated EISCAT heater signal is relatively near the observed count, the increase of the observed signal is quite similar to the increase of the transmitted signal. During the period No 8, instead, no clear increase of the signal was detected. Moreover, during the period No 6 the maximum signal strength was much less than on its neighboring period Nos 4, 5 and 7. Note that the comparison between the observations and the simulated signal is also limited because of the ~ 280 data point gaps during the period Nos 5 and 9. It is also unknown if the narrow data peak between period Nos 4 and 5 is a real signal or is it an artefact associated with the data gap.

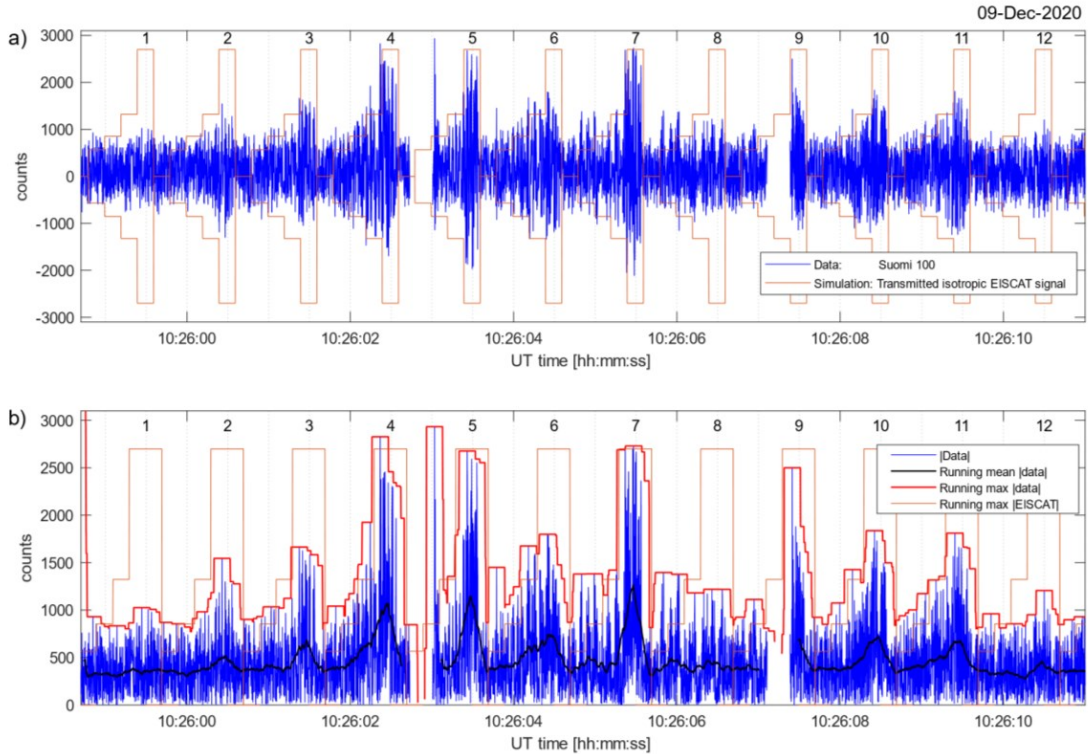

**Figure 7**. Comparison between observations of the HEARER instrument on the Suomi 100 satellite during the flyby on Dec. 9, 2020, and the synthetic EISCAT heater signal. a) The blue line shows the observed signal. The orange lines mimic the behavior of the electric field of the unattenuated transmitted signal where the power of the transmitted EISCAT heater signal increases at every 0.2 second as 0%, 4.4%, 10%, 24% and 100% of its maximum power. b) The blue line shows the absolute value of the observed signal and the black line its 200-point running mean. The red and the orange line shows a 200-point running maximum of the absolute value of the observations and the absolute value of the simulated EISCAT heater signal, respectively. In panels a) and b) the observations are divided into 12 one-second time periods, the numbers of which are shown on the top of the figure. Note that there is a data gap in the data between period Nos 4–5 and within No 9 caused by the instrument's onboard data operations.

Note that the synthetic EISCAT heater signal shown in Fig. 7 did not contain antenna pattern because the goal of the short time period and, consequently specially localized, comparison was to compare the periodicity of the observed and transmitted signal. An important question, however, arises: how may the transmitting antenna pattern have affected the intensity of the observed signal? This was investigated in Fig. 8a which shows the amplification (GL) of the antenna pattern of the EISCAT heater's Array 1 as well as the path of the orbit of the Suomi 100 satellite on Dec. 9, 2020. As can be seen in Fig. 8a, the amplification depends strongly on the azimuthal and elevation angle. The amplification is strongly peaked around the zenith and the satellite passes over several local minimum and maximum amplification directions when the elevation angle is positive, i.e., when the satellite is over the horizon viewed from the EISCAT heater.

The simulated intensity of the EISCAT heater signal along the orbit of the Suomi 100 satellite during its flyby on Dec. 9, 2020, is shown in Fig. 8b. The electric field, $E$, is derived by interpolating the amplitude shown in Fig. 8a along the orbit of the spacecraft (the while line). The intensity of the signal ($I$[W/m$^2$]) is then obtained from the used maximum transmission power $P_{max}$[W], which in the analyzed case was $7.55 \times 10^5$ W, and the distance between the spacecraft and the EISCAT heater, $r$[m], as $I = P_{max} \, 10^{GL/10} / (4 \pi r^2)$ [W/m$^2$]. Although the key parameter which determined the signal in the ferrite rod antenna is the magnetic field ($B$), or more precisely according to Faraday's law, its time variations, it is informative to derive the value of the electric field because radio waves are often measured by electric field instruments. The electric field along the orbit was derived by approximating a vacuum condition, for simplicity, when $I = E \times B / \mu_o = E^2 / (c \, \mu_o)$, where c is the speed of light in a vacuum and $\mu_o$ is the vacuum magnetic permeability. Attenuation was approximated to be zero because, as already mentioned, the attenuation by absorption in the analyzed case is only about 1.3% and the signal was detected at the closest distance. This slightly underestimates the drop of the signal from its peak value because the ionosphere decreases slightly the intensity in the analyzed time range.

As can be seen from Fig. 8b, there is a local minimum in the signal strength around its global maximum about 15 seconds before and after the main beam's peak value. Therefore, as seen in Fig. 8b (the black dashed line), the signal weakens much faster along the orbit than it would decrease if the signal had been transmitted isotopically. In that case, the intensity would have decreased as $1/r^2$ and, consequently, the electric field in vacuum conditions as $1/r$, as illustrated in Fig. 8b. However, it is anticipated that the HEARER will observe a strongly peaked narrow signal because of the EISCAT heater's antenna pattern. Indeed, this was observed to be the case on Dec. 9, 2020 (Fig. 8c). Therefore, the comparison between the simulated heater signal (Fig. 8b) and the observed signal (Fig. 8c) suggests that the observed signal is narrow and decreases rapidly from the observed peak because of the antenna pattern of the EISCAT heater.

It is worth noting that in addition to this general trend of the observed signal, variations with the signal can result from factors other than the EISCAT antenna pattern. Especially, detailed interpretation of the origin of the signal observed by the HEARER instrument is complicated by the fact that the radiation pattern of a small ferrite-rod antenna is omnidirectional where the theoretical radiation pattern of an electrically small antenna is $sin(\theta)$, where $\theta$ is the angle between the direction of the arriving signal and the axis of the rod. Consequently, the gain is null only for a signal which arrives along the axis of the rod antenna. This effect cannot, however, be taken into account in the signal strength analysis presented in this paper, because the attitude of the satellite and, therefore, the ferrite-rod antennas, remains unknown. The receiver's antenna pattern is, however, not expected to alter significantly the received signal strength during a single flyby because of the relatively small total observation time of the signal compared with the rotation period of the satellite, which is typically about 2 minutes. In such a situation, the direction of the HEARER's antenna remains relatively unchanged during the time when the signal was observed. One should note that a non-optimal antenna orientation, that may have coincidentally occurred during some of the flybys, might explain why we fail to detect the signals in these flybys. This remains, however, a speculative point. It should be finally noted that the ionosphere can also have small-scale 3D electron density structures which can change the intensity of the observed signal, as seen earlier in Fig. 3 and as will be discussed later in Section 5.

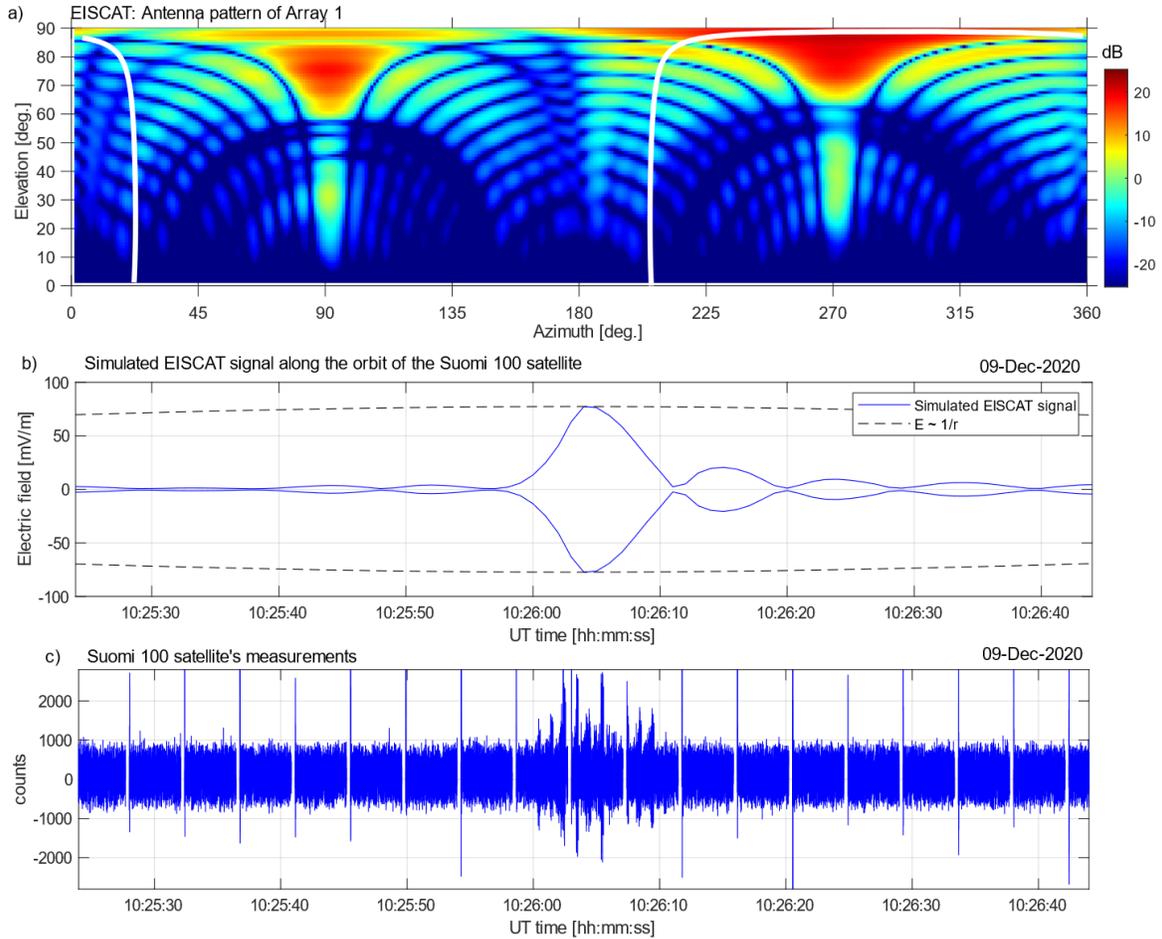

**Figure 8**. a) Amplification of the signal transmitted from the EISCAT heater Array No 1 at 600 km from the heater at different elevation and azimuthal angles. The color shows the amplification (G) in $10\times\log_{10}$ and its maximum value is 21.7. The white line shows the elevation and azimuth of the Suomi 100 satellite during its flyby on Dec. 9., 2020. The satellite raises above the horizon at the azimuthal angle ~ 20°, and moves below the horizon at the azimuthal angle ~ 205°. b) (blue line) Simulated peak electric field along the satellite's orbit which is shown in Fig. 8a. Both the magnitude and its opposite number is shown because the transmitted signal is a sinusoidal wave. The decrease in the electric field from the peak value as one over the distance from the EISCAT heater is shown for a comparison (dashed black line). c) The observed signal in the same time range on Dec. 9, 2020.

### 4.1 Quantitative analysis of the HEARER' measurements

The relatively small amount of the data which includes the clearest EISCAT Heater signal, i.e., about 10 s interval on Dec. 9 challenges the possibility to perform a comprehensive quantitative data analysis. However, already the obtained measurements provide (i) valuable information about the sensitivity of the HEARER instrument and (ii) observations can also be used to investigate possible non-linear effects associated with the heating of the ionosphere with the EISCAT heater.

First, a comparison of the simulated intensities and observed counts showed in Figs. 8b and 8c provides a possibility to make in-space calibration of the HEARER instrument. This is important additional usage of the heating campaign because the detailed noise level caused by the satellite in the measured frequency range as well as the sensitivity of the HEARER instruments were unknown before the launch. This is because the ferrite-rod antennas on the satellite's mock-up frame were tested using Aalto University's gigahertz transverse electromagnetic (GTEM) cell, but these antenna tests were made without the satellite. Moreover, no electromagnetic compatibility (EMC) testing of the satellite was possible to be made before the launch because of the strict time schedule of the Suomi 100 satellite project.

Fig. 8b showed that the peak observed signal, i.e., the signal associated with the EISCAT Heater, is about 1600 counts above the average peak of the background. Because the simulated peak electric field along the orbit is about 77 mV m$^{-1}$, then 1 count corresponds to the electric field of about 0.05 mV m$^{-1}$ (~ 77 mV m$^{-1}$ / 1600 counts) in the used frequency. In addition to this information concerning the resolution of measurements, one can also give estimation for the minimum possible EISCAT heater signal strength what can be detected by the HEARER instrument. The estimation was made by manually adding a similar type of "1s / 0.2 s" intensity ramp shape into the observed background data, and by increasing the peak value of the added signal until it can be identified unambiguously from the background. Estimation by visual inspection indicated that about a 250 count maximum value can still be observed unambiguously. This means that the signal would have been detected even if the strength of the observed electric field would have been about 250/1600 = 0.16 times smaller than it was on Dec 9, 2020, i.e., approximately 77 mV m$^{-1}$ × (250/1600) ~ 12 mV/m strong electric field.

### 4.1.1 A non-linear "self-absorption" detected

While passaging through the plasma, a powerful radio wave can generate a variety of nonlinear effects in which the propagation path of the wave is altered by the radio wave's energy (Gurevich, 2007). In the following, the nonlinear relationship between the transmitted and received power of our experiment is studied. In the lower panel of Fig. 7 at modulation 8, we saw a clear reduction in the received power from the linear trend when the ERP in the satellite direction is expected to be at its highest (i.e., at modulations 4–9). This nonlinearity is expected to result from the effect known as "self-action", or "self-absorption", in which the increased transmitter power is not only increasing the reception at the receiver, but also heats the ionospheric plasma along the propagation path, which in turn reduces the ionospheric opacity for the radio waves (e.g., Gurevich, 2007).

Here we have estimated the magnitude of this nonlinearity by calculating the collisional Joule heating and its impact on the ionospheric transparency (see Kero et al., 2000, for details of the used method). As a model for the radio wave propagation, the generalized Appleton theory (Sen and Wyller, 1960) was used and the collection of cooling mechanisms were given by Stubbe and Varnum (1972). The electron density profile was the same as shown in Fig. 3 and the neutral atmospheric composition and temperature were given by the MSISE-model (Hedin, 1991).

The transmitter power (755 kW) and the antenna gain (27.1 dB at maximum), were adopted from the EISCAT log-readings recorded during the experiment. These logs also indicate that, in addition to the O-mode transmission, a significant portion (up to 28%) of the X-mode wave was also sent during the experiment. Hence, the actual heating effect was likely something between the ones modelled for both the O- and the X-mode. As a result, the heating model predicts the maximum heating effect in the deep D-region, around 70 km in altitude, where the electron temperature is increased by a factor of 3.4 (O-mode) or 4.0 (X-mode) (Fig. 9 left panel). The equivalent maximum reductions in intensity above the D region were 28% for the O-mode and 43% for the X-mode (Fig. 9 right panel).

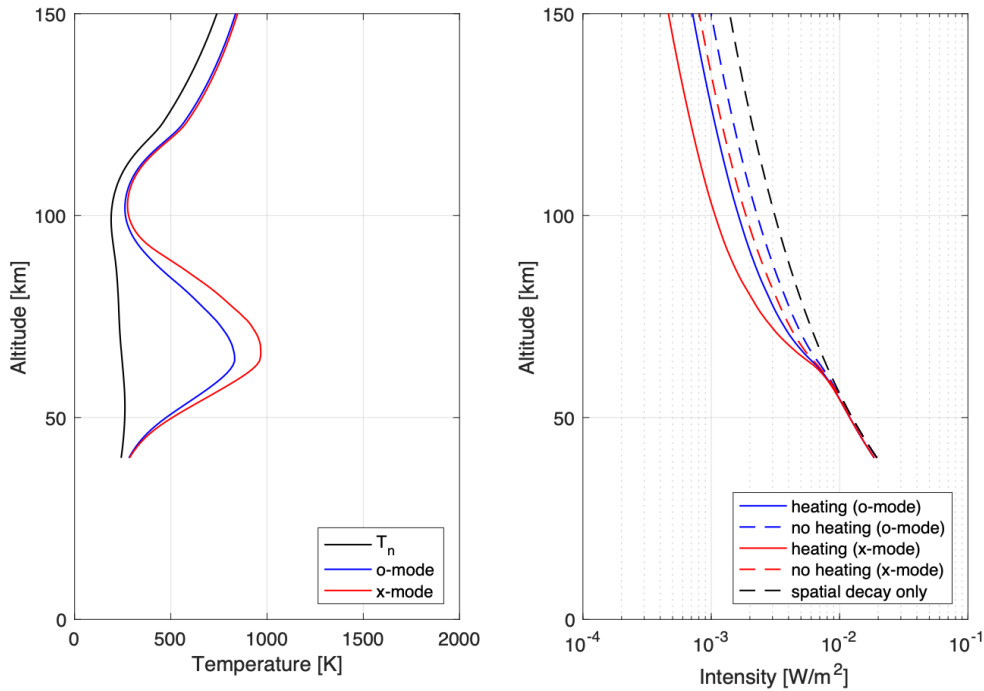

**Figure 9**. a) Heating modelling for the Suomi 100 flyby on Dec. 9., 2020, i.e., based on the electron density profile show in Fig. 3. Left panel: the expected electron heating at the maximum power (here 755 kW) and in the center of the Heater beam main lobe (antenna gain 27.2 dBi). Red line: O-mode, blue line: X-mode and black: the background neutral temperature. Right panel: intensity of the radio wave. Black dashed line represents the spatial decay only, i.e., the intensity without any ionospheric losses. The colored dashed lines (blue for O-mode, red for X-mode) show the intensity of the radio wave absorbed by the plasma that is in the neutral background temperature, i.e., no heating effects are taken into account. The solid lines (blue for O-mode, red for X-mode) show the intensities affected by the electron heating shown in the left panel.

The modelled self-absorption effect is studied against the Suomi 100 data in Fig. 10. The comparison of the normalized powers of the EISCAT Heater and the Suomi 100 observations shown in the lower panel reveal a clear reduction in the received signal compared to the linear model, although the modulation cycles show remarkable variation is power. As can be seen in Fig. 10b, modulations 1-3, similarly to 10–12 (during small to moderate heating) follow roughly the linear response. Modulations with strong heating 5-8 show that, in general, a remarkable reduction in the received power, compared to the linear response, is observed, although not monotonically as a function of heating. Modulations close to the maximum heating agree well with the model predictions (30–40% nonlinearity), while the Suomi 100 reception at the center of the Heater beam (modulation 6) shows an unexpectedly large power reduction (data: ~80%, model: 27-38%), but still a clearly detectable Heater signal.

In contrast, during the modulation 8, the Suomi 100 satellite does not receive any recognizable heater signal, and the reason for this remains unclear. There are no visible artefacts in the Suomi 100 data and the Heater power logging does not reveal any power cut-offs, although a short anomaly might be lost between its 5-s sampling intervals. In any case, the modulation 8, with no visible signature of the Heating signal, can be probably ignored as an outlier. The moderate heating at modulations 4 and 9 seem to produce a mixed result: 4 shows no significant heating effect, while 9 shows a strong reduction of 55%.

As a summary, a significant "self-absorption" of the HF signal power is observed with respect to the heating power. The observed relative non-linearity ranges from 30–40%, i.e., the range expected by the Heating model, to the unexpectedly large 80% reduction observed at the center of the heater beam. This paper is, to author's best knowledge, the first observation of this kind of "self-absorption" measured from the transionospheric signal path from a powerful radio source on the ground a satellite-borne receiver.

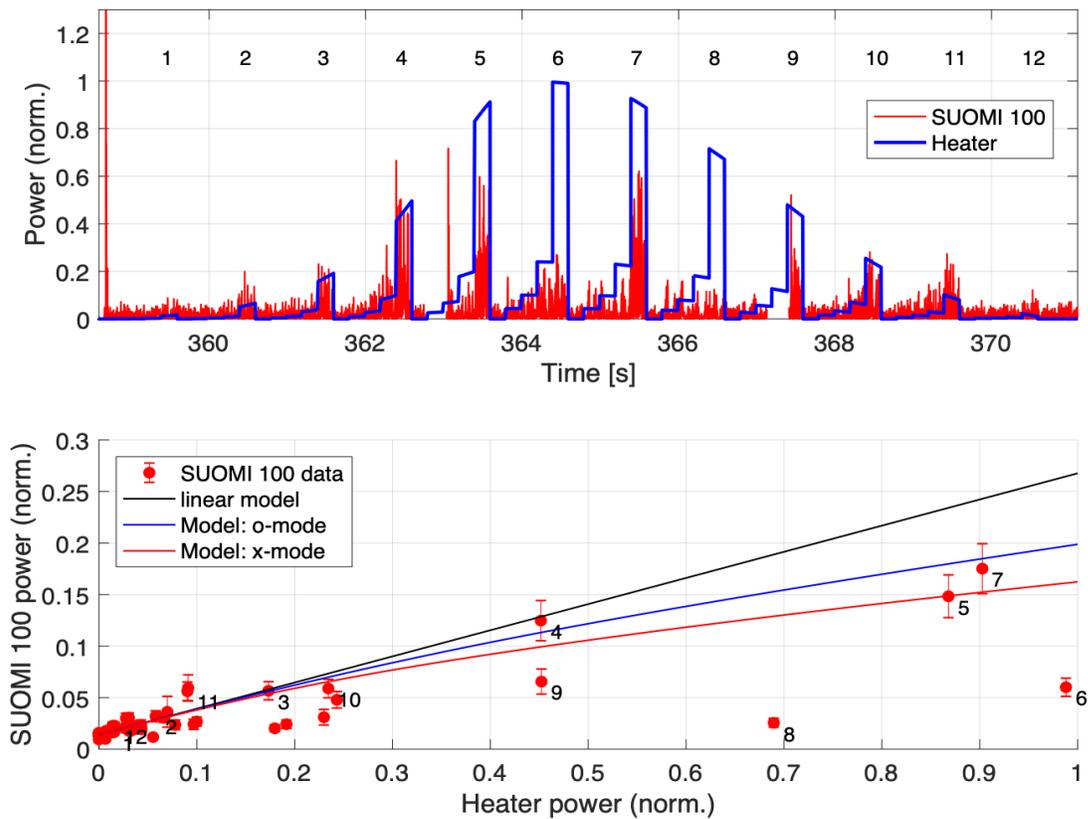

**Figure 10**. The Suomi 100 received power vs. the Heater power. Upper panel: The Suomi 100 normalized received power (red), vs. the Heater power modulation (blue) multiplied by the Heater antenna gain along the satellite orbit (Fig. 8b). The normalized Suomi 100 power is defined as the received amplitude squared (see Fig. 8c), normalized so that the power is $1.2 \times 10^7 \times$ (count$^2$). Note, that the somewhat arbitrary normalization value does not affect any conclusions based on the relative values, and the normalization was chosen to follow roughly the heater modulation (blue line). The Heater power modulations are numbered 1–12 at the times of the maximum power, consistently with Fig. 7. Lower panel: Suomi 100 power vs. the Heater power averaged over each 0.2-s modulation periods (red dots with 2-sigma-error bars). Heating model results (see Fig. 9) are shown as a blue and a red line, for the O and the X mode, respectively. The black line represents a linear model response between the Suomi 100 satellite and the Heater obtained by fitting to the data points for which the Heater power is less than 10% of the maximum (i.e., Heater power (norm.) is less than 0.1). For these power levels, the heating effect is expected to be insignificant. Note also that the Suomi 100 power peak at the beginning of the first modulation is ignored as an outlier.

## 5 Discussion

This paper presents the first results from the joint measurements where the EISCAT Heater facility transmitted a HF wave, which was received by the HEARER instrument on the Suomi 100 satellite. Three out of 14 satellite measurements in 2020-2021 showed a transmitted signal from EISCAT, on Dec. 8 noon and evening flybys and the clearest signal on the Dec. 9 noon flyby in 2020. In this paper the clearest signal was analyzed in detail.

The reason why the signal was detected only during three of the campaign flybys needs further detailed investigation in the future. Differences may be associated with different seasons or different local times. The properties of the ionosphere may also have been different than in the following days. Moreover, different frequencies (7.0 MHz and 7.953 MHz) and different modes (the X-mode and the O-mode) were also used in the campaigns. In addition, all EISCAT's transmitter antennas were not always used at full power during the 2-hour heating periods. On the other hand, during the 2019 measurement campaign, the signal was not detected probably because at that time the EISCAT heater used antenna Array 3 which has a significantly smaller maximum ERP power in the used frequency (7.953 MHz) than Array 1 which was used during the 2020 campaign. The direction of HEARER's antenna may also have been unfavorable to observe the signal on some passes. However, a detailed comparison between all flybys is beyond the scope of the present study.

The analysis of the Dec. 9, 2020, flyby demonstrated how the source of the signals observed by the HEARER type of radio spectrometers can be interpreted with a ray-tracing simulation. The presented ray-tracing analysis can be, however, regarded to be only the first step to utilize the observations, because the ray tracing analysis was made using a 1D electron density altitude profile based on the EISCAT IS radar and digisonde observations everywhere for simplicity. Moreover, the IRI model itself provides smooth changes in the electron density profile without large electron density gradients. As seen from the EISCAT VHF measurements of the electron densities below 200 km, in practice, the ionosphere has small-scale structures (c.f. Fig. 3). Small-scale electron density enhancements, in turn, cause scintillation of radio waves (e.g., Basu et al., 1998; van der Meeren, 2015; James et al., 2017). The EISCAT heater also heats the ionosphere resulting in plasma irregularities (see e.g., Yampolski et al., 2019, and references therein) and additional scintillation (Bernhardt et al., 2016). These phenomena may also affect the changes in the observed signal strengths because another possibility for the variability of the HF signal amplitudes at the satellites, apart from self-attenuation, is focusing and fading of the waves by refraction on HF-induced irregularities. Consequently, in some positions interference will amplify and other positions weaken the signals. Satellite observations in the HF range can also be used to improve ionosphere models, such as FORTE measurements to the IRI model (Moses and Jacobson, 2004). Ionospheric densities can also be estimated by using Global Positioning System (GPS) signals combined with mathematical models, e.g., such as has been used to make ionospheric tomography by using GPS signals with the TomoScand algorithm (see e.g., Nordberg et al., 2018). In the future, when there are more HEARER measurements available, one can use these measurements to investigate how modifying of the IRI electron density profiles affects the simulated intensities along the orbit of the satellite.

The presented analysis provided several lessons on the technology point of view of how to increase the usage of small-satellite radio-instrument measurement and observation campaigns. First, the attitude of the radio instrument's antenna, and therefore the attitude of the spacecraft, is important information when the observations are analyzed in detail. In the presented analysis, that information was unfortunately not available, but the attitude information can be derived in the future either by collecting housekeeping data or by using the camera. In the presented analysis these possibilities were not used during the measurements in order to minimize electromagnetic noise from the satellite.

The second lesson is that, if the radio instrument is not calibrated in detail on ground, signals from a ground-based transmitter can be used to make in-orbit calibration of the radio instrument. Making calibration on ground is, of course, beneficial because calibration in space requires modelling where all inputs are not known in detail, such as the ionospheric density values and probably the attitude and detailed position of the spacecraft. Interestingly, this kind of in-orbit calibration can be made, and has also been made, for the Suomi 100's camera instrument (Knuuttila et al., 2022).

From the scientific point of view, the presented campaign can also be regarded as being the first step towards a comprehensive joint satellite-radar campaigns. In all measurements the transmitted frequency was either a constant 7.0 MHz or 7.953 MHz. Scientifically more rich measurements may, however, be obtained when several frequencies are used during the flyby, because the ionosphere affects differing frequencies differently and provides a range of information about the ionosphere. For example, the making of 0.2 s ramps 0.0–0.2 s: OFF, 0.2–0.4 s: 3.5 MHz, 0.4–0.6 s: 5 MHz, 0.6–0.8 s: 6.5 MHz and 0.8–1.0 s: 8 MHz, could be anticipated to be scientifically highly interesting. Valuable new information may also be obtained by using both the O-mode and the X-mode waves during a flyby. When the intensity of the transmitted signal is changed in time, as done in the cases analyzed in this paper, one also has an opportunity to investigate possible non-linear effects in the ionosphere in greater detail than shown in Fig. 10. In the future, using EISCAT_3D observations (McCrea et al., 2015) would make it possible to extend the analysis shown in Fig. 3 for measured volumetric electron density profiles from D to F region altitudes.

The practical disadvantage of such lower frequency measurements with the EISCAT heater would be that, as already mentioned, the maximum ERP of all the Heaters's arrays decrease approximately linearly with decreasing frequency. This is because the lower the frequency, the wider the beam and the weaker the intensity. On the other hand, the presented analysis implied that one would have observed the EISCAT heater signal even if the signal had been about 20% of the signal used on Dec. 9, 2020.

It should be finally noted that the analysis presented in this paper was based on the observations made with a single small spacecraft. The disadvantage of using such small spacecraft is that, because the necessary optimization in terms of the mass, size, power consumption and data handling of the possible instrumentation affects the ultimate quality of the data, the data analysis of the data can be relatively demanding. On the other hand, the relative inexpensive nanosatellites may open a door for a satellite constellation, where the same signal is observed simultaneously by several radio instruments, which can be used with the aid of tomography to provide valuable information about the 3D ionosphere and its dynamics.

# 6 Conclusions

This paper presented the first results of measurement with the HEARER radio instrument on-board the Suomi 100 satellite, the first Finnish space research satellite. We presented, to the authors' knowledge, the first ever successful active satellite-radar transionospheric measurements in the HF range with a CubeSat-sized satellite. These observations are unique because they were made with a small 1 Unit CubeSat and a simple radio spectrometer.

The 7.953 MHz signal transmitted from the EISCAT HF facility was observed by the satellite when it was near the transmitter. Interpretation of the observations made on Dec. 9, 2020, showed that the changes in the observed signal associated with the transmitter's antenna pattern and shape of the transmitted signal are clearly identified and, consequently, these changes can be separated from the effects associated with the properties of the ionosphere. To the author's knowledge, the work contains the first observation of "self-absorption" measured from the transionospheric signal path from a powerful radio source on the ground to a satellite-borne receiver.

The analysis also suggested that the active satellite-heater measurement campaigns in the future are anticipated to provide even more useful information about the ionosphere than the presented cases if several frequencies are used in a flyby, and not only a single frequency, as used in the analyzed measurements. Overall, the study demonstrates that transionospheric HF propagation observations can be successfully made already with a small, relatively simple and lightweight radio spectrometer on a small nanosatellite.

## Acknowledgments

The authors would like to thank Finnish Prime Minister's Office and Magnus Ehrnrooth Foundation for the financial support of the Suomi 100 satellite project. The work of AK is funded by the Tenure Track Project in Radio Science at Sodankylä Geophysical Observatory/University of Oulu. EK thanks Jakke Mäkelä, Niko Porjo, Juha Mallat and Jari Hänninen for useful discussions about narrow-band radio receivers and Markku Mäkelä for mechanical and electronic help. EK and JN thank Clemens Icheln for the information about Aalto University's GTEM device. Heikki Vanhamäki, Esa Turunen and Tuija Pulkkinen are acknowledged for the invaluable discussions about the usage of radio waves in ionosphere research and for the support of the Suomi 100 satellite project. EK thanks Hanna Rothkaehl and Tomasz Szewczyk for the discussions about small-sized radio spectrometers. EK also thanks Jari Mäkinen for the Suomi 100 satellite outreach activities and Valtteri Harmainen for the design and implementing of the data-sharing cloud service. EK and JN thank Rami Stenman for on-ground HF-calibration support at Akaa, Finland. The authors thanks the EISCAT community for the heater measurement time and for its operations. Michael Rietveld is acknowledged for valuable suggestions and remarks about the manuscript. The authors also thank the IRI community for the electron density model.

**Open Research**

The analyzed HEARER instrument observations will be available on request.

**References**

Bahcivan, H., & Cutler, J. W. (2012). Radio Aurora Explorer: Mission science and radar system, *Radio Science*, 47, RS2012. https://doi.org/10.1029/2011RS004817

Bahcivan, H., Cutler, J. W., Bennett, M., Kempke, B., Springmann, J. C., Buonocore, J., et al. (2012). First measurements of radar coherence scatter by the radio aurora explorer cubeSat, *Geophys. Res. Lett.*, 39, 1–5. https://doi.org/10.1029/2012GL052249

Bahcivan, H., Cutler, J. W., Springmann, J. C., Doe, R., & Nicolls, M. J. (2014). Magnetic aspect sensitivity of high-latitude E region irregularities measured by the RAX-2 CubeSat, *J. Geophys. Res. Space Physics*, 119, 1233–1249. https://doi.org/doi:10.1002/2013JA019547

Basu, S., Weber, E. J., Bullet, T. W., Keskinen, M. J., MacKenzie, E., Doherty P., Sheehan, R., Kuenzler, H., Ning, P., & Bongiolatti, J. (1998). Characteristics of plasma structuring in the cusp/cleft region at Svalbard, *Radio Science*, 33(6), 1885–1899. https://doi.org/doi:10.1029/98RS01597

Bernhardt, P. A., Siefring, C. L., Briczinski, S. J., McCarrick, M., & Michell R. G. (2016). Large ionospheric disturbances produced by the HAARP HF facility, *Radio Science*, 51, 1081–1093. https://doi.org/10.1002/2015RS005883

Budden K.G. (1985). The Propagation of Radio Waves: The Theory of Radio Waves of Low Power in the Ionosphere and Magnetosphere, Cambridge University Press.

Dokukin, V. S. (1992). "The APEX project scientific facility orbital complex," in APEX Project. Scientific Purposes, Simulation, Technique and Equipment of Experiment, eds V. N. Oraevsky and Y. Y. Ruzhin (Moscow: Nauka), 16–29. In Russian.

Fish C.S. et al. (2014). Performance of the Dynamic Ionosphere CubeSat Experiment Mission, *Space Sci Rev.*, 181, pp. 61–120. https://doi.org/10.1007/s11214-014-0034-x

Fontell, M. (2019). Numerical Ray Tracing of Medium and High Frequency Radio Waves in the Terrestrial Ionosphere, MSc thesis, School of Electrical Engineering, Aalto University. http://urn.fi/URN:NBN:fi:aalto-201902031505

Franklin, C. A., & Maclean, M. A. (1969). The design of Swept-frequency topside sounders, Proc. IEEE, 57, 897–929. https://doi.org/10.1109/PROC.1969.7135

Gurevich, A. V. (2007). Nonlinear effects in the ionosphere. *Phys. Usp.*, 50, 1091–1121. https://doi.org/10.1070/PU2007v050n11ABEH006212

Hedin, A. E. (1991). Extension of the MSIS Thermospheric Model into the Middle and Lower Atmosphere, *J. Geophys. Res.*, 96, 1159. https://doi.org/10.1029/90JA02125

Holden, D. N., Munson, C. P., & Devenport, J. C. (1995). Satellite observations of transionospheric pulse pairs, *Geophys. Res. Lett.*, 22(8), 889-892. https://doi.org/10.1029/95GL00432

James, H. G., Inan, U. S., & Rietveld, M. T. (1990). Observations on the DE 1 spacecraft of ELF/VLF waves generated by an ionospheric heater, *J. Geophys. Res.*, 95, A8, 12,187-12,195. https://doi.org/10.1029/JA095iA08p12187

James. H. G. (2006). Effects on transionospheric HF propagation observed by ISIS at middle and auroral latitudes, *Advances in Space Research*, 38, 2303–2312. https://doi.org/10.1016/j.asr.2005.03.114

James H. G., King, E.P., White, A., Hum, R.H., Lunscher, W. H. H. L., & Siefring, C.L. (2015). The e-POP Radio Receiver Instrument on CASSIOPE, *Space Sci Rev*, 189, 79–105. https://doi.org/10.1007/s11214-014-0130-y

James, H. G., Frolov, V. L., Andreeva, E. S., Padokhin, A. M., & Siefring, C. L. (2017). Sura heating facility transmissions to the CASSIOPE/e-POP satellite, *Radio Science*, 52, 259–270. https://doi.org/10.1002/2016RS006190

Karpachev A. T., Klimenko, M. V., Klimenko, V. V., & Kuleshova, V. P. (2013). Statistical study of the F3 layer characteristics retrieved from Intercosmos-19 satellite data, *Journal of Atmospheric and Solar-Terrestrial Physics*, 103, 121–128. https://doi.org/10.1016/j.jastp.2013.01.010

Kero, A., Bösinger, T., Pollari, P., Turunen, E., & Rietveld, M. (2000). First EISCAT measurement of electron-gas temperature in the artificially heated D-region ionosphere, *Ann. Geophys.*, 18, 1210–1215. https://doi.org/10.1007/s00585-000-1210-8

King J. W. (1963). Investigations of the upper ionosphere deduced from top-side sounder data, *Nature*, 197, 639 – 641. https://doi.org/10.1038/197639a0

Knuuttila, O., Kallio, E., Partamies, N., Syrjäsuo, M., Kauristie, K., Sofieva, V., Alho A., et al. (2022). In-space Calibration of Nanosatellite Camera, *The Journal of Small Satellites*. *submitted*.

Kopf, A. J., Gurnett, D. A., Morgan, D. D., & Kirchner, D. L. (2008). Transient layers in the topside ionosphere of Mars, *Geophys. Res. Lett.*, 35, L17102. https://doi.org/doi:10.1029/2008GL034948

Koskimaa, P. (2016). Ferrite Rod Antenna in a Nanosatellite Medium and High Frequency Radio, MSc thesis, School of Electrical Engineering, Aalto University. http://urn.fi/URN:NBN:fi:aalto-201611025371

Leyser, T. B., James, H. G., Gustavsson, B., & Rietveld, M. T. (2018). Evidence of L-mode electromagnetic wave pumping of ionospheric plasma near geomagnetic zenith, *Ann. Geophys.*, 36, 243-251. https://doi.org/10.5194/angeo-36-243-2018

Massey, R. S., Knox, S. O., Franz, R. C., Holden, D. N., & Rhodes, C. T. (1998). Measurements of transionospheric radio propagation parameters using the FORTE satellite, *Radio Science*, Volume 33, Number 6, 1739-1753. https://doi.org/10.1029/98RS02032

McCrea, I, Aikio, A., Alfonsi, L., Belova, E., Buchert, S., Clilverd M., et al. (2015). The Science case for the EISCAT_3D radar, *Progress in Earth and Planetary Science*, 2, 21. https://doi.org/10.1186/s40645-015-0051-8

Meftah, M., Boust, F., Keckhut, P., Sarkissian, A., Boutéraon, T., Bekki, S., et al. (2022). INSPIRE-SAT 7, a Second CubeSat to Measure the Earth's Energy Budget and to Probe the Ionosphere. *Remote Sens.*, 14, 186. https://doi.org/10.3390/rs14010186

Moses R. W., & Jacobson, A. R. (2004). Ionospheric profiling through radio-frequency signals recorded by the FORTE´ satellite, with comparison to the International Reference Ionosphere, *Advances in Space Research*, 34, 2096–2103. https://doi.org/10.1016/j.asr.2004.02.018

Norberg, J., Vierinen, J., Roininen, L., Orispää, M., Kauristie, K., Rideout, W. C, et al. (2018). Gaussian Markov Random Field Priors in Ionospheric 3-D Multi-instrument Tomography, *IEEE Transactions of Geoscience and Remote Sensing*. https://doi.org/10.1109/TGRS.2018.2847026

Oraevsky, V. N., Pulinets, S. A., Bud'ko, N. I., Prutensky, I. S., Vaskov, V. V., Klos, Z., et al. (1998). Emissions stimulated in the upper ionosphere by the SURA heating facility, *Adv. Space Res.*, 21, 671–675. https://doi.org/10.1016/S0273-1177(97)01002-8

Perry, G. W., James, H. G., Gillies, R. G., Howarth, A., Hussey, G. C., McWilliams, K. A., et al., (2017). First results of HF radio science with e-POP RRI and SuperDARN, *Radio Science*, 52, 78–93. https://doi.org/doi:10.1002/2016RS006142

Perry, G. W., Frissell, N. A., Miller, E. S., Moses, M., Shovkoplyas, A., Howarth, A. D., & Yau, A. W. (2018). Citizen radio science: An analysis of amateur radio transmissions with e-POP RRI. *Radio Science*, 53, 933–947. https://doi.org/10.1029/2017RS006496

Priedhorsky W. C., Bloch, J. J., Wallin, S. P., Armstrong, W. T., Siegmund, O. H. W., Griffee, J., Fleete, R., (1993). The ALEXIS Small Satellite Project: Better, Faster, Cheaper Faces Reality, *IEEE Transactions On Nuclear Science*, 40, 4. https://doi.org/10.1109/NSSMIC.1992.301362

Poghosyan, A., & Golkar, A. (2017). CubeSat evolution: Analyzing CubeSat capabilities for conducting science missions, *Progress in Aerospace Sciences*, 88, 59-83. https://doi.org/10.1016/j.paerosci.2016.11.002

Prech, L, Ruzhin, Y.Y., Dokukin, V.S., Nemecek, & Z., Safrankova, J., (2018). Overview of APEX Project Results, *Front. Astron. Space Sci.*, 5:46. https://doi.org/10.3389/fspas.2018.00046

Rietveld, M. T., Senior, A., Markkanen, J., & Westman, A. (2016). New capabilities of the upgraded EISCAT high-power HF facility, *Radio Science*, 51, 1533–1546. https://doi.org/10.1002/2016RS006093

Rietveld, M. T., & Stubbe, P. (2021). History of the Tromsø Ionosphere Heating facility, *Hist. Geo Space. Sci.* Discuss. [preprint]. https://doi.org/10.5194/hgss-2021-19, under review

Rothkaehl, H., & Klos, Z. (1996). HF radio emissions as a tool of ionospheric plasma diagnostic, *Annali Di Geofisica*, Vol. XXXIX, 4.

Sen, H. K., & Wyller, A. A. (1960), On the generalization of the Appleton-Hartree magnetoionic formulas, *J. Geophys. Res.*, 65( 12), 3931– 3950. https://doi.org/10.1029/JZ065i012p03931

Streltsov, A. V., Berthelier, J.-J., Chernyshov, A. A., Frolov, V. L., Honary, F., Kosch, M. J., et al. (2018). Past, Present and Future of Active Radio Frequency, Experiments in Space, *Space Sci Rev.*, 214:118. https://doi.org/10.1007/s11214-018-0549-7

Stubbe, P., & Varnum, W. S. (1972). Electron energy transfer rates in the ionosphere, *Planet. Space Sci.*, 20, 1121-1126. https://doi.org/10.1016/0032-0633(72)90001-3

Teodossiev, D. K., Astrukova, M. S., Shkevov, R. G., & Galev G. K. (2001). Low-frequency electric field measurement probes on board the Intercosmos-24 AKTIVEN satellite, *Aerospace Research in Bulgaria* (ISSN 0861-1432), No. 16, pp. 46 – 53.

Tjulin, A., (2017). EISCAT Experiments, EISCAT Scientific Association. https://eiscat.se/wp-content/uploads/2017/04/Experiments.pdf

van der Meeren, C., Oksavik, K., Lorentzen, D. A., Rietveld, M. T. & Clausen, L. B. N. (2015). Severe and localized GNSS scintillation at the poleward edge of the nightside auroral oval during intense substorm aurora, *J. Geophys. Res. Space Physics*, 120. https://doi.org/10.1002/2015JA021819

Vas'kov, V. V., Bud'ko, N. I., Kapustina, O. V., Mikhailov, Yu. M., Ryabova, N.A., Gdalevich, G.L., Komrakov, G.P., & Maresovc, A.N. (1998). Detection on the Intercosmos-24 satellite of VLF and ELF waves stimulated in the topside ionosphere by the heating facility Sura, *Journal of Atmospheric and Solar Terrestrial Physics*, 60, 1261-1274. https://doi.org/10.1016/S1364-6826(98)00054-6

Yampolski Y, Milikh, G., Zalizovski, A., Koloskov, A., Reznichenko, A., Nossa, E. et al. (2019) Ionospheric Non-linear Effects Observed During Very-Long-Distance HF Propagation, *Front. Astron. Space Sci.*, 6:12. https://doi.org/10.3389/fspas.2019.00012

Winter L. M. & Ledbetter K. (2015). Type II and type III radio bursts and their correlation with solar energetic proton events, *The Astrophysical Journal*, Volume 809, Number 1. https://doi.org/10.1088/0004-637X/809/1/105

Zabotin, N. A., Zavorotny, V. U. & Rietveld M. T. (2014). Physical mechanisms associated with long-range propagation of the signals from ionospheric heating experiments, *Radio Science*, 49, 987–995. https://doi.org/10.1002/2014RS005573

Zhang, X., Frolov, V., Zhou, C., Zhao, S., Ruzhin, Y., Shen, X., et al. (2016). Plasma perturbations HF-induced in the topside ionosphere, *J. Geophys. Res.*, 121, 10052–10063. https://doi.org/10.1002/2016JA022484